\RequirePackage{fix-cm}
\documentclass[smallcondensed]{svjour3}     
\smartqed  
\usepackage{graphicx}
\usepackage[us,12hr]{datetime}

\begin{document}

\title{STE-QUEST Mission and System Design}
\subtitle{Overview after completion of Phase-A}
\titlerunning{STE-QUEST Mission and System Design}        

\author{Gerald Hechenblaikner \and
        Marc-Peter Hess \and
        Marianna Vitelli \and
        Jan Beck
}
\institute{G. Hechenblaikner, M.-P. Hess, M. Vitelli, J. Beck, \at
              EADS Astrium, 88039 Friedrichshafen, Germany\\
              \email{gerald.hechenblaikner[at]astrium.eads.net} \\
}

\maketitle

\begin{abstract}
STE-QUEST is a fundamental science mission which is considered for launch within the Cosmic Vision programme of the European Space Agency (ESA).
Its main scientific objectives relate to probing various aspects of Einstein's theory of general relativity by measuring the gravitational red-shift of the earth, the moon and the sun as well as testing the weak equivalence principle to unprecedented accuracy. In order to perform the measurements, the system features a spacecraft equipped with two complex instruments, an atomic clock and an atom interferometer, a ground-segment encompassing several ground-terminals collocated with the best available ground atomic clocks, and clock comparison between space and ground via microwave and optical links. The baseline orbit is highly eccentric and exhibits strong variations of incident solar flux, which poses challenges for thermal and power subsystems in addition to the difficulties encountered by precise-orbit-determination at high altitudes. The mission assessment and definition phase (Phase-A) has recently been completed and this paper gives a concise overview over some system level results.
\end{abstract}

\section{Introduction and Mission Objectives}
\label{intro}
Space Time Explorer \& Quantum Equivalence Space Test (STE-QUEST) is a medium-sized mission candidate for launch in 2022/2024 in the Cosmic Vision programme of the European Space Agency. After recommendation by the Space Science Advisory Committee, it was selected to be first studied by ESA, followed by two parallel industrial assessment studies. This paper gives a brief summary of the assessment activities by Astrium which build on and extend the preceding ESA study as described in\cite{gehler2013esa}.

STE-QUEST aims to study the cornerstones of Einstein's Equivalence Principle (EEP), pushing the limits of measurement accuracy by several orders of magnitude compared to what is currently achievable in ground based experiments\cite{schiller2010space,cacciapuoti2013ste-quest}.
On the one hand, experiments are performed to measure the gravitational red-shift experienced by highly accurate clocks in the gravitational fields of earth or sun (Space Time Explorer). On the other hand, differential accelerations of microscopic quantum particles are measured to test the universality of free fall, also referred to as Weak Equivalence Principle (Quantum Equivalence Space Test).
These measurements aim at finding possible deviations from predictions of General Relativity (GR), as postulated by many theories trying to combine GR with quantum theory. Examples include deviations predicted by string theory \cite{damour1994string}, loop quantum gravity\cite{mattingly2005modern}, standard model extension\cite{colladay1998lorentz}, anomalous spin-coupling\cite{moody1984new}, and space-time-fluctuations\cite{goklu2008metric}, among others.
The STE-QUEST mission goal is summarized by the four primary science objectives\cite{cacciapuoti2013ste-quest} which are listed in Tab.\ref{tab:mission_objectives} together with the 4 measurement types geared at achieving them.
\begin{table}
\caption{The STE-QUEST mission objectives and measurement types.}
\label{tab:mission_objectives}       
\begin{tabular}{lll}
\hline\noalign{\smallskip}
Primary Mission Objective & Measurement Accuracy & Measurement Strategy  \\
\noalign{\smallskip}\hline\hline\noalign{\smallskip}
{\bf Objective 1:} Measurement of & to a fractional frequency  & (a) Space-clock comparison  \\
Earth gravitational red-shift &  uncertainty better than  & to ground clock at apogee\\
&$1\times 10^{-7}$& (b) Space-clock comparison \\
&& between apogee and perigee\\
\noalign{\smallskip}\hline\noalign{\smallskip}
{\bf Objective 2:} Measurement of &  to a fractional frequency & (c) Comparison between \\
Sun gravitational red-shift &  uncertainty better than  & two ground clocks via\\
&$2\times10^{-6}$, goal: $5\times 10^{-7}$& spacecraft\\
\noalign{\smallskip}\hline\noalign{\smallskip}
{\bf Objective 3:} Measurement of &  to a fractional frequency & (c) Comparison between \\
Moon gravitational red-shift &  uncertainty better than  & two ground clocks via\\
&$4\times10^{-4}$, goal: $9\times 10^{-5}$& spacecraft\\
\noalign{\smallskip}\hline\noalign{\smallskip}
{\bf Objective 4:} Measurement of & to an uncertainty in the  & (d) Atom Interferometer \\
Weak Equivalence Principle & E\"otv\"os param. smaller & measurements at perigee\\
&$1.5\times 10^{-15}$&\\
\noalign{\smallskip}\hline\noalign{\smallskip}
\end{tabular}
\end{table}

\section{Mission Overview}
\begin{figure}
\begin{center}
\includegraphics[width=1.0\textwidth]{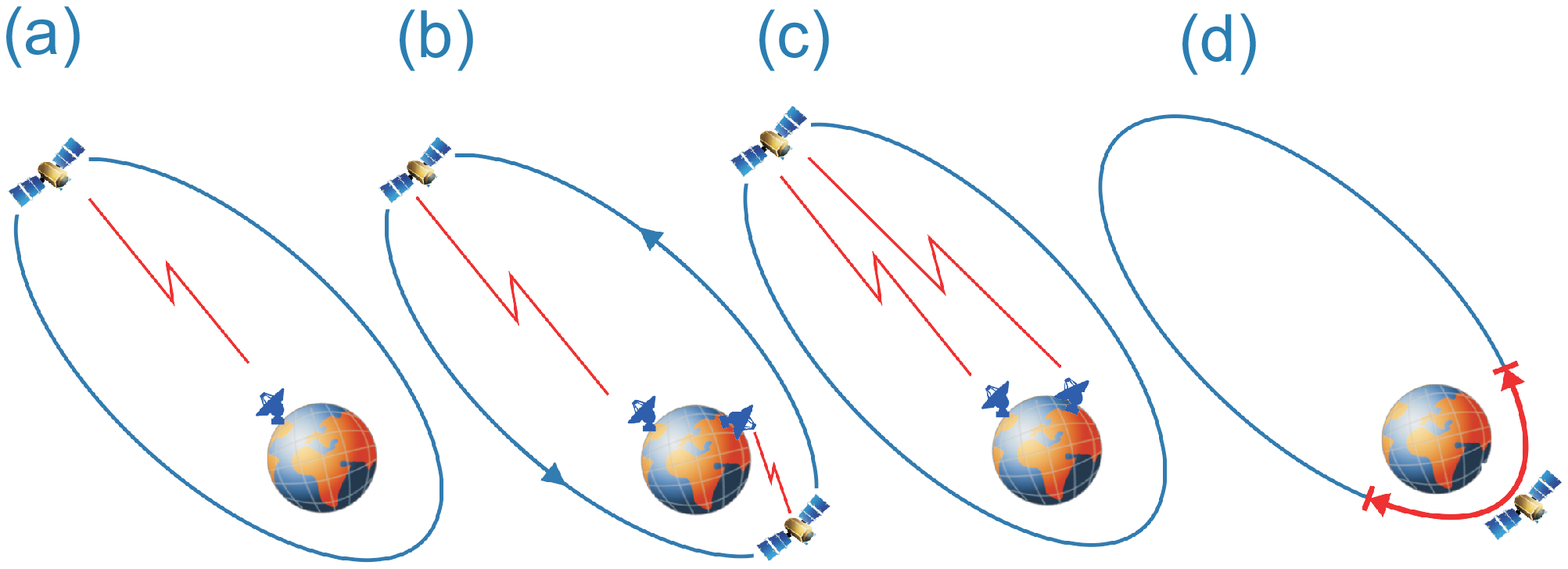}
\caption{A schematic of the 4 STE-QUEST measurement types as described in Tab.\ref{tab:mission_objectives}}
\label{fig:measurement_principle}       
\end{center}
\end{figure}

The STE-QUEST mission is complex and comprises a space-segment as well as a ground segment, with both contributing to the science performance. Highly stable bi-directional microwave (X/Ka-band) and optical links (based on laser communication terminals) connect the two segments and allow precise time-and-frequency transfer. The space-segment encompasses the satellite, the two instruments, the science link equipment, and the precise orbit determination equipment. The ground-segment is composed of 3 ground terminals that are connected to highly accurate ground-clocks.
In order to fulfil the mission objectives, various measurement types are used that are shown in Fig.\ref{fig:measurement_principle}. We shall briefly discuss them.
\subsection{Measurement Types}
{\it Earth gravitational red-shift measurements:} The frequency output of the on-board clock is compared to that of the ground clocks. In order to maximize the signal, i.e. the relativistic frequency offset between the two clocks, a Highly Elliptical Orbit (HEO) is chosen. When the spacecraft is close to earth during perigee passage, there are large frequency shifts of the space-clock due to the strong gravitational field. When it is far from earth during apogee passage, there are only small gravitational fields and therefore small frequency shifts.
Whilst measurement type (a) compares the space-clock at apogee to the ground-clock, relying on space-clock accuracy, measurement type (b) compares the frequency variation of the space-clock over the elliptical orbit, which requires clock stability in addition to visibility of a ground terminal at perigee (see also section \ref{sec:orbit_ground}).
\\[1em]
{\it Sun gravitational red-shift measurements:} The relativistic frequency offset between a pair of ground clocks is compared while the spacecraft merely serves as a link node between the ground clocks for a period of typically 5 to 7 hours, as described by measurement type (c). In that case, the accuracy of the space-clock is of minor importance whereas link stability and performance over a long period are essential. As this requirement is also hard to fulfil by the microwave links alone, the optical links play an important role. However, availability and performance of optical links are strongly affected by weather conditions which in turn depend on the location and altitude of the ground terminals.
\\[1em]
{\it Moon gravitational red-shift measurements:} As in the preceding case, the relativistic frequency offset between a pair of ground clocks is compared while the spacecraft merely serves as a link node between the ground clocks. The potential signals for a violation of the EEP with the moon as the source mass can be easily distinguished from those with the sun as the source mass due to the difference in frequency and phase.  It is important to point out that unless the EEP is violated the measured frequency difference between two distant ground clocks (due to the sun or the moon) is expected to be zero up to small tidal corrections and a constant offset term from the earth field\cite{studyteam2013yellowbook}.
\\[1em]
{\it Atom interferometer measurements:} These measurements do not require contact to ground terminals but must be performed in close proximity to earth where the gravity gradients are large and the E\"otv\"os parameter (see Eq.\ref{equ:otvos} below) becomes small. The latter defines the magnitude of a possible violation of the weak equivalence principle (WEP), as required for mission objective 4. For this reason, the atom interferometer operates during perigee passage at altitudes below 3000 km, where gravity accelerations exceed $4.5~{\rm m/s^2}$.

\subsection {Payload}
The STE-QUEST payload primarily consists of two instruments, the atomic clock and the atom interferometer, as well as equipment for the science links (microwave and optical) and for precise orbit determination (POD). Whilst a detailed description of the proposed payload elements is found in \cite{schiller2010space,studyteam2013ste-quest,studyteam2013yellowbook}, here we shall only give a brief overview over the payload as summarized in Table \ref{tab:payload}. Note that some payload elements, such as the optical links and the onboard clock, may become optional in a future implementation of STE-QUEST, which is discussed in the official ESA assessment study report (yellow book)\cite{studyteam2013yellowbook}.
\begin{table}
\caption{Elements of the STE-QUEST payload.}
\label{tab:payload}       
\begin{tabular}{lll}
\hline\noalign{\smallskip}
Payload element, & Subsystem, & Components  \\
Instrument & Unit &   \\
\noalign{\smallskip}\hline\hline\noalign{\smallskip}
{\bf Instrument 1:}   & PHARAO NG & caesium tube, laser source \\
{\bf Atomic Clock} &&\\
& Microwave-Optical Local   & laser head, frequency comb, \\
& Oscillator (MOLO)& highly-stable cavity\\
&&\\
& MW Synthesis \& Frequency & ultra-stable oscillator, Frequency Gen.,\\
& Distribution (MSD)& Comparison and Distribution (FGCD) \\
&&\\
&Instr. Control Unit (ICU)& control electronics, harness\\
\noalign{\smallskip}\hline\noalign{\smallskip}
{\bf Instrument 2:}  &  Physics Package & vacuum chamber, atom source,  \\
{\bf Atom Interf.} && magnetic coils and chip, laser \\
&&interfaces, mu-metal shields\\
&&\\
& Laser Subsystem  & cooling, repumping, Raman, \\
&& detection, optical trapping lasers\\
&&\\
&Electronics & ion pump control, DMU, magnetic coil \\
&& drive, laser control electronics\\
\noalign{\smallskip}\hline\noalign{\smallskip}
{\bf Science Link}  &  Microwave links & MW electronics, X-/Ka-band antennas \\
{\bf Equipment} &&\\
& Optical links  & 2 Laser Communication Terminals \\
&&(LCTs), synchronization units\\
\noalign{\smallskip}\hline\noalign{\smallskip}
{\bf POD} & GNSS equipment & GNSS receiver and antennas \\
{\bf Equipment} && \\
 & Laser ranging equipment  & corner cube reflectors (optional)\\
\noalign{\smallskip}\hline\noalign{\smallskip}
\end{tabular}
\end{table}
\\[1em]
{\it The atomic clock} is essentially a rebuilt of the high-precision PHARAO clock used in the ACES mission\cite{salomon2001cold,cacciapuoti2009space}, with the additional provision of an optically derived microwave signal (MW) for superb short term stability which is generated by the MOLO. In PHARAO, a cloud of cold caesium atoms is prepared in a specific hyperfine state and launched across a vacuum tube. There, the cloud is interrogated by microwave radiation in two spatially separated Ramsey zones before state transition probabilities are obtained from fluorescence detection. Repeating the cycle while scanning the MW fields yields a Ramsey fringe pattern whose period scales inversely proportional to the flight time T between Ramsey interactions.

In the MOLO, a reference laser is locked to an ultra-stable cavity, which provides superb short-term stability in addition to the long-term stability and accuracy obtained from PHARAO by locking the reference signal to the hyperfine transition of the caesium atoms.
The reference laser output is frequency-shifted to correct for long term drifts and then fed to the femtosecond frequency comb\cite{udem2002optical,ye2005femtosecond} to convert the signal from the optical to the microwave-domain.
Core components of the MOLO, such as the highly stable cavity\cite{webster2011force,folkner2010laser,legero2010tuning}, the frequency comb\cite{kozma2007metrology,plattner2010optical}, and optical fibres \cite{lezius2012radiation} are currently being developed or tested in dedicated national programs. Additionally, other components of optical clocks have been investigated in the ongoing Space-Optical-Clock program (SOC)\cite{schiller2007optical} of the European Space Agency.
\\[1em]
{\it The atom interferometer}\cite{kasevich1991atomic,rasel1995atom} uses interference between matter waves to generate an interference pattern that depends on the external forces acting on the atoms under test. In order to reject unwanted external noise sources as much as possible, differential measurements are performed simultaneously with two isotopes of Rubidium atoms, namely ${\rm ^{87}Rb}$ and ${\rm ^{85}Rb}$. The measured phases for both isotopes are subtracted to reveal any possible differences in acceleration under free fall conditions of the particles. This way, the common mode noise which is present on both signals is rejected to a high degree\cite{schubert2013differential,aguilera2013ste}.
In the proposed instrument \cite{schubert2013differential,aguilera2013ste} a number of the order of $10^6$ atoms of both isotopes are trapped and cooled by means of sophisticated magnetic traps in combination with laser cooling. Thus, the residual temperature and consequently the residual motion and expansion of the atomic clouds is reduced to the nK regime, where the cloud of individual atoms makes the transition to the Bose-Einstein condensed state. The atoms are then released from the trap into free fall before being prepared in superpositions of internal states and split into partial waves by pulses of lasers in a well defined time sequence.
By using a sequence of 3 laser pulses separated by a period of $5\,{\rm s}$, a Mach-Zehnder interferometer in space-time is formed. The distribution of the internal atomic states is then detected either by fluorescence spectroscopy or by absorption imaging. The distribution ratio of the internal atomic states provides the information about the phase shift which is proportional to the acceleration acting on the atoms.
The E\"otv\"os parameter $\eta$ (see mission objective 4) depends on the ratio of inertial mass $m_i$ to gravitational mass $m_g$ for the two isotopes and can be found from the measured accelerations $a$ as follows:
\begin{equation}
\label{equ:otvos}
\eta=2\frac{(m_g/m_i)_{1}-(m_g/m_i)_{2}}{(m_g/m_i)_{1}+(m_g/m_i)_{2}}=2\frac{a_{1}-a_{2}}{a_{1}+a_{2}},
\end{equation}
where subscript 1 refers to physical parameters of the first isotope (${\rm ^{87}Rb}$) and subscript 2 to those of the second isotope (${\rm ^{85}Rb}$).
\\The programmatics of European space atom interferometers, as proposed for STE-QUEST strongly builds on the different national and European-wide activities in this direction, in particular the Space Atom Interferometer (SAI)\cite{sorrentino2010compact} and the nationally funded projects QUANTUS\cite{van2010bose} and ICE\cite{stern2009light}. In these activities innovative technologies are developed and tested in parabolic flights, drop tower campaigns and sounding rocket missions.
\\[1em]
{\it The science links} comprise optical and microwave time \& frequency links to compare the on-board clock with distributed ground clocks. Similarly, the links allow to compare distant ground clocks to one another using the on-board segment as a relay.
The microwave link (MWL) is designed based on re-use and further development of existing MWL technology for the ACES mission\cite{hess2011aces,seidel2007aces}, while implementing lessons learned from ACES. This applies to the flight segment and the ground terminals, as well as for science data post-processing issues down to the principles and concept for deploying and operating a network of ground terminals and associated ground clocks. Digital tracking loops with digital phase-time readout are used to avoid beat-note concepts. Different to ACES and given the highly eccentric orbit, Ka-band is used instead of Ku-band, X-band instead of S-band for ionosphere mitigation, and the modulation rate is changed from 100 MChip/s to 250 MChip/s. The MWL flight segment comprises four receive channels to simultaneously support comparisons with 3 ground terminals and on-board calibration.

The optical link baseline design relies on two TESAT Laser Communication Terminals (LCTs)\cite{gregory2010tesat} on-board the satellite, both referenced to the space-clock MOLO signal, and the ground LCTs to perform two time bi-directional comparisons. The LCTs operate at 1064 nm with a 5 W fiber laser, similar to the EDRS LCT\cite{heine2011laser}, and a 250 MHz RF modulation. A telescope aperture of 150 mm in space and 300 mm on ground was chosen. A preliminary performance estimate has been made by extrapolation from existing measurements with more detailed investigations following now in dedicated link studies. The optical links are found to be particularly sensitive to atmospheric disturbances and local weather conditions. These particularly affect common view comparisons between two ground terminals which are typically located in less favorable positions concerning local cloud coverage.
\\[1em]
{\it The Precise-Orbit-Determination (POD) equipment} is required to determine the spacecraft position and velocity, which allows calculating theoretical predictions of the expected gravitational red-shifts and the relativistic Doppler, and compare them to the measured ones. Our studies found that a GNSS-receiver with a nadir-pointing antenna suffices to achieve the required accuracy\cite{hechenblaikner2013gnss}. Additional corner-cube reflectors to support laser ranging measurements, such as also used for the GIOVE-A satellite\cite{urschl2008orbit}, may optionally be included as a backup and for validation of GNSS-derived results.

\subsection{Orbit and ground terminals}
\label{sec:orbit_ground}
\begin{figure}
\begin{center}
\includegraphics[width=1.0\textwidth]{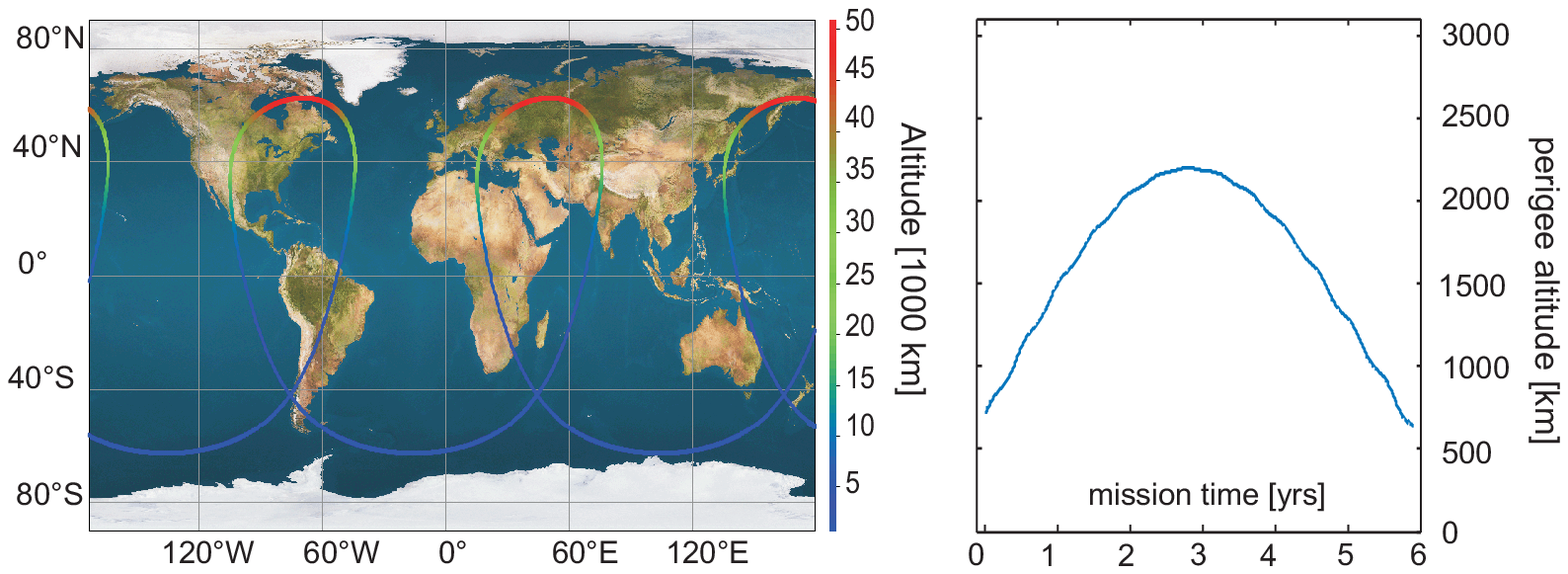}
\caption{(Left) The ground-track of the STE-QUEST baseline orbit features a 3:2 resonance. (Right) The perigee altitude is plotted against mission time.}
\label{fig:orbit}       
\end{center}
\end{figure}

{\it The baseline orbit} is highly elliptical with a typical perigee altitude of 700 km and an apogee altitude of 50000 km, which maximizes gravity gradients between apogee and perigee in order to achieve mission objective 1 (earth gravitational red-shift). The corresponding orbit period is 16h, leading to a 3:2 resonance of the ground track as depicted in Fig.\ref{fig:orbit} left. Some useful orbit parameters are summarized in Tab. \ref{tab:baseline_orbit}.
\begin{table}
\caption{Key parameters for the STE-QUEST baseline orbit.}
\label{tab:baseline_orbit}
\begin{tabular}{llllll}
\hline\noalign{\smallskip}
perigee alt. & apogee alt. & orbit period & inclination & arg. perigee & RAAN drift\\
\noalign{\smallskip}\hline\noalign{\smallskip}
700-2200 km & $\sim$ 50 000 km&16~h& $62.6^{\circ}$& $271^{\circ}$ & $0.105^{\circ}$\\
\noalign{\smallskip}\hline\noalign{\smallskip}
\end{tabular}
\end{table}
Contrary to the old baseline orbit\cite{renk2012ste}, where the argument of perigee was drifting from the North to the South during the mission, the argument of perigee for the new baseline orbit is frozen in the South. Therefore, contact to the ground terminals located in the Northern hemisphere can only be established for altitudes higher than 6000 km, which considerably impairs the modulation measurements of the earth gravitational red-shift (measurement type b in Table\ref{tab:mission_objectives}). On the pro side, for higher altitudes and in particular around apogee, common view links to several terminals and over many hours can be achieved which greatly benefits the sun gravitational red-shift measurements (type c).
The baseline orbit also features a slight drift of the right-ascension of the ascending node (RAAN) due to the $J2$ term of the non-spherical earth gravitational potential\cite{montenbruck2000satellite}, which leads to a total drift of approximately 30$^{\circ}$ per year. This drift being quite small, the orbit can be considered as nearly-inertial during one year so that seasonal variations of the incident solar flux are expected (see also Fig.\ref{fig:solar_flux}).

The major advantage of the new baseline orbit over the old one is a greatly reduced  $\Delta$V for de-orbiting, which allows reducing the corresponding propellant mass from 190 kg to 60 kg. The reason for that is the evolution of the perigee altitude due to the 3-body interaction with the Sun and the Moon\cite{montenbruck2000satellite}. The perigee altitude starts low at around 700 km, goes through a maximum of around 2200 km at half time, and then drops back down below 700 km in the final phase, as depicted in Fig.\ref{fig:orbit} right. The low altitude at the end of mission facilitates de-orbiting using the 6 x 22-N OCT thrusters during an apogee maneuver that is fully compliant with safety regulations.
\\[1em]
{\it The Ground segment} comprises 3 ground terminals, each equipped with microwave antennas for X- and Ka-band communication as well as (optional) optical terminals based on the Laser Communication Terminals (LCT) of Tesat\cite{gregory2010tesat}. The terminals must be distributed over the earth surface with a required minimum separation of five thousand kilometers between them. This  maximizes their relative potential difference in the gravitational field of the sun in pursuit of mission objective 2. Additionally, the terminals must be in the vicinity of high-performance ground clocks (distance $<100$ km) to which they may be linked via a calibrated glass-fibre network. This narrows the choice of terminal sites to the locations of leading institutes in the field of time-and frequency metrology. Factoring in future improvements in clock technology and design, those institutes are likely able to provide clocks with the required performance of a fractional frequency uncertainty on the order of $10^{-18}$) at the start of the mission, considering that current clock performance comes already quite close to the requirement\cite{hinkley2013atomic,bloom2013a,rosenband2008frequency}.
Accounting for all the aspects mentioned above, Boulder (USA), Torino (Italy), and Tokyo (Japan) were chosen as the baseline terminal locations.

\subsection{Spacecraft Overview}
\begin{figure}
\begin{center}
\includegraphics[width=1.0\textwidth]{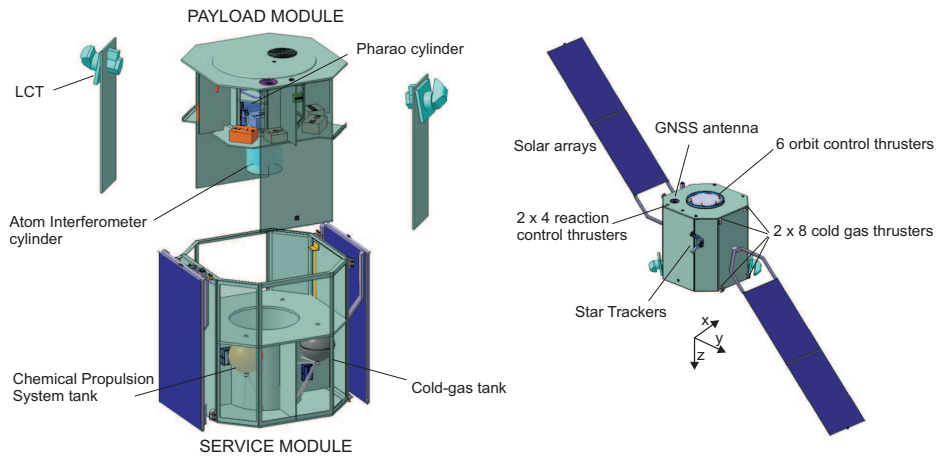}
\caption{(Left) STE-QUEST spacecraft with stowed solar arrays, separating into PLM and SVM. (Right) Spacecraft thruster configurations and deployed solar arrays.}
\label{fig:spacecraft_overview}       
\end{center}
\end{figure}

The spacecraft has an octagonal shape with a diameter of 2.5 m and a height of 2.7. It consists of a payload module (PLM), accommodating the instruments as well as science link equipment, and a service module (SVM) for the spacecraft equipment and propellant tanks (see Fig. \ref{fig:spacecraft_overview}).
The Laser Communication Terminals (LCTs) are accommodated towards the nadir end of opposite skew panels. The nadir panel (+z-face) accommodates a GNSS antenna and a pair of X-band/Ka-band antennas for the science links. The telemetry \& tele-commanding (TT\&C) X-band low gain and medium gain antennas are accommodated on y-panels of the service module. The two solar array wings are canted at $45^{\circ}$ with respect to the spacecraft y-faces and allow rotations around the y-axis. Radiators are primarily placed on the +x/-x faces of the spacecraft and good thermal conductivity between dissipating units and radiators is provided through the use of heat-pipes which are embedded in the instrument base-plates.
A set of 2 x 4 reaction control thrusters (RCTs) for orbit maintenance and slew maneuvers after the perigee measurement phase is used as part of the chemical propulsion system (CPS) together with 1 x 6 Orbit Control thrusters (OCTs) for de-orbiting.  The latter is becoming a stringent requirement for future mission designs to avoid uncontrolled growth of space debris and reduce the risk of potential collisions between debris and intact spacecraft\cite{janovsky2004end,klinkrad2004esa}. The Micro-Propulsion System (MPS) is based on GAIA heritage\cite{matticaricold} and primarily used for attitude control, employing a set of 2x8 micro-proportional thrusters to this purpose. As primary sensors of the attitude and orbit control subsystem (AOCS), the spacecraft accommodates 3 star trackers, 1 inertial measurement unit (IMU), and 6 coarse sun sensors.

\begin{figure}
\begin{center}
\includegraphics[width=0.75\textwidth]{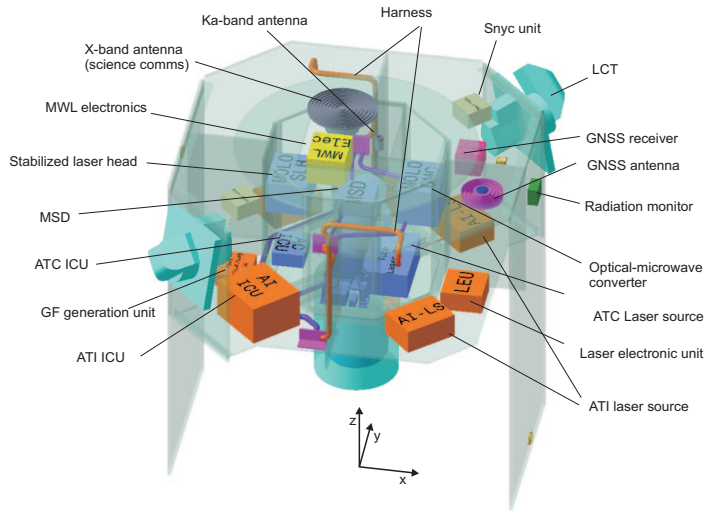}
\caption{The Payload Module with accommodated instrument units.}
\label{fig:payload_module}       
\end{center}
\end{figure}

The payload module of Fig. \ref{fig:payload_module} shows how the two instruments are accommodated in the well protected central region of the spacecraft, which places them close to the center-of-mass (CoM) and therefore minimizes rotational accelerations whilst also providing optimal shielding against the massive doses of radiation accumulated over the five years of mission duration.
The PHARAO tube of the atomic clock is aligned with the y-axis. This arrangement minimizes the Coriolis force which acts on the cold atomic cloud propagating along the vacuum tube axis when the spacecraft performs rotations around the y-axis. These rotations allow preserving a nadir-pointing attitude of the spacecraft as required for most of the orbit (see also section \ref{sec:aocs}). The sensitive axis of the atom interferometer must point in the direction of the external gravity gradient. It is therefore aligned with the z-axis (nadir) and suspended from the spacecraft middle plate into the structural cylinder of the service module which provides additional shielding against radiation and temperature variations.

The total wet mass of the spacecraft with the launch adapter is approximately 2300 kg, including unit maturity margins and a 20\% system margin. The total propellant mass (CPS+MPS) is 260 kg, out of which 60 kg are reserved for de-orbiting.  The total power dissipation for the spacecraft is 2230 W, including all required margins, if all instruments and payload units operate at full power. The spacecraft is designed to be compatible with all requirements (including launch mass, envelope, loads, and structural frequencies) for launch with a Soyuz-Fregat from the main ESA spaceport in Kourou.

\section{Performance}
\subsection{Attitude and Orbit Control}
\label{sec:aocs}
\begin{figure}
\begin{center}
\includegraphics[width=1.00\textwidth]{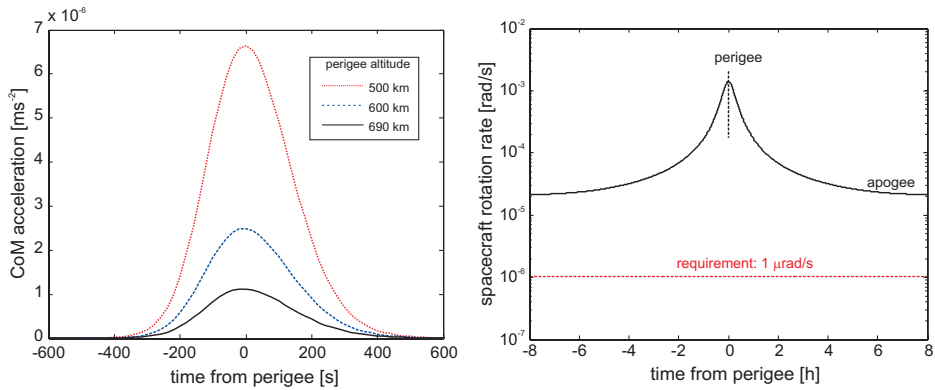}
\caption{(Left) The drag acceleration acting on the spacecraft is plotted against the time interval from perigee for various perigee altitudes. (Right) The spacecraft rotation rate is plotted against the time interval from perigee, assuming a nadir-pointing strategy. The maximum tolerable level for atom interferometry is given by the red broken line. }
\label{fig:drag_and_rotation}       
\end{center}
\end{figure}

The two instruments pose stringent requirements on attitude stability and non-gravitational accelerations which drive the design of the attitude and orbit control system (AOCS) and the related spacecraft pointing strategy.
To ensure full measurement performance, the atom interferometer requires non- gravitational center-of-mass accelerations to be below a level of $10^{-6}\,{\rm m/s^{2}}$. Among the external perturbations acting on the spacecraft and causing such accelerations, air drag is by far the dominant one. It scales proportional to the atmospheric density and therefore decreases with increasing altitudes. Figure \ref{fig:drag_and_rotation} (left) displays the level of drag accelerations for various altitudes and under worst case assumptions, i.e. maximal level of solar activity and maximal cross section offered by the spacecraft along its trajectory. The results indicate that below altitudes of approximately 700 km drag accelerations become intolerably large.
We conclude that, for the baseline orbit, drag forces are generally below the required maximum level and therefore need not be compensated with the on-board micro-propulsion system, although such an option is feasible in principle.

\begin{figure}
\begin{center}
\includegraphics[width=1.00\textwidth]{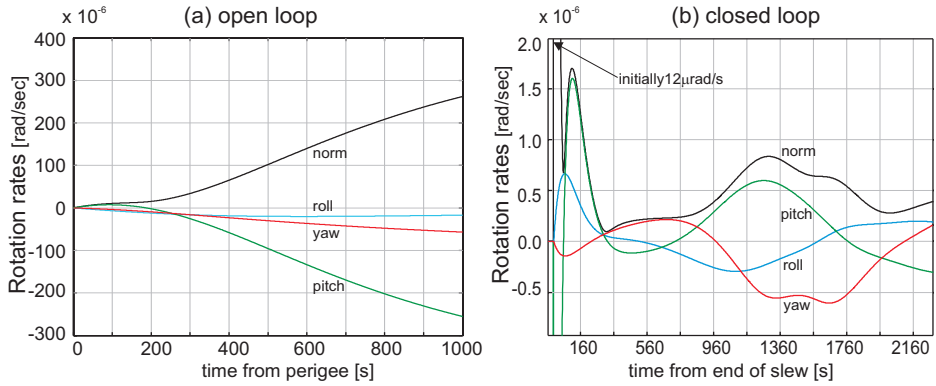}
\caption{Spacecraft rotation rates for roll, pitch, yaw angles and the norm during perigee passage. (Left) The spacecraft is freely floating, starting from perigee (t=0) with assumed initial rates of zero. (Right) The spacecraft is under closed loop control which is activated right after the slew maneuver (t=0), a few hundred seconds before entering perigee passage. Note that the high initial pitch rate of 12 $\mu{\rm rad/s}$ is outside the plot scale.}
\label{fig:rate_errors}
\end{center}
\end{figure}

Another requirement is given by the maximal rotation rate of $10^{-6}\,{\rm rad/s}$ during perigee passage when sensitive atom interferometry experiments are performed. Figure \ref{fig:drag_and_rotation} (right) plots the spacecraft rotation rate (black solid line) against the time for one entire orbit, assuming a nadir-pointing strategy. The plot demonstrates that the requirement (red broken line) would not only be violated by the fast rotation rates encountered at perigee, even for the comparatively slow dynamics at apogee the rotation rates would exceed the requirement by more than one order of magnitude. We therefore conclude that atom interferometer measurements are incompatible with a nadir-pointing strategy throughout the orbit, requiring the spacecraft to remain fully inertial during such experiments. Considering that atom interferometry is generally performed at altitudes below 3000 km, the following spacecraft pointing strategy was introduced for STE-QUEST (see Figure \ref{fig:pointing_strategy}):
\begin{figure}
\begin{center}
\includegraphics[width=1.00\textwidth]{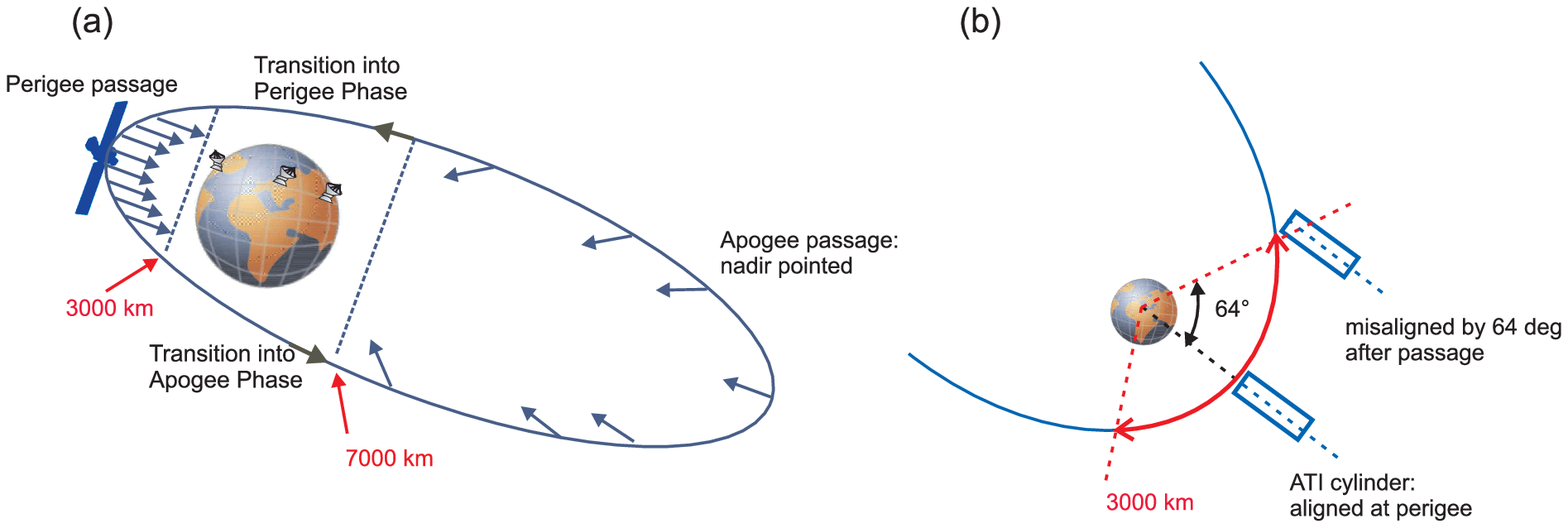}
\caption{(Left) The STE-QUEST spacecraft pointing strategy with its 3 phases. (Right) During atom interferometry the ATI-cylinder is aligned with the gravity gradient at perigee.}
\label{fig:pointing_strategy}       
\end{center}
\end{figure}
\begin{itemize}
\item {\bf Perigee Passage:} For altitudes below 3000 km, lasting about $1900\,{\rm s}$ each orbit, the atom interferometer is operating and the spacecraft remains inertial. The atom interferometer axis is aligned with nadir at perigee and rotated away from nadir by 64 degrees at 3000 km altitude (see Figure \ref{fig:pointing_strategy}). The solar array driving mechanism is frozen to avoid additional micro-vibrations.
\item {\bf Transition into apogee:} For altitudes between 3000 km and 7000 km, right after the perigee passage, the spacecraft rotates such that it is nadir pointed  again (slew maneuver). The solar arrays are unfrozen and rotated into optimal view of the sun.
 \item	{\bf Apogee Passage:} For altitudes above 7000 km the spacecraft is nadir pointed and rotational accelerations are generally much slower than during perigee passage. The solar array is allowed to rotate continuously for optimal alignment with the sun. Atom interferometer measurements are generally not performed.
 \item {\bf Transition into perigee:} For altitudes between 7000 km and 3000 km, right after the apogee passage, the spacecraft rotates such that it is aligned with nadir at perigee. The solar arrays are frozen.
\end{itemize}
During the science operations at perigee and apogee, the attitude control relies solely on the MPS. The slew maneuvers, i.e. $64^{\circ}$ pitch rotations, are performed by means of the CPS in $400\,{\rm s}$ at a rate of ${\rm 0.26\,deg/s}$. The CPS is also used to damp out residual rotational rates which are caused by solar array bending modes and fuel sloshing due to the fast maneuvers. When the perigee passage is entered, the MPS is activated to further attenuate the residual rate errors coming from the minimum impulse bit and sensor bias (ca. ${\rm 10^{-5}\,rad/s}$), which is shown in Fig.\ref{fig:rate_errors}b. Closed-loop simulations, based on a dynamical computer model of the spacecraft in-orbit and simulated AOCS equipment, demonstrated that the damping requires about 500 s. The complete transition, including maneuvers, lasts about ${\rm 900\, s}$. In order to mitigate the effects of external disturbances during the perigee passage, an AOCS closed-loop control has been designed that is based on cold-gas thrusters and includes roll-off filtering to suppress sensor noise while maintaining sufficient stability margins. This allows compensating external torques from drag forces, solar radiation pressure and the earth magnetic field (among others), yielding rotation rates well below the required $10^{-6}\,{\rm rad/s}$ after the initial damping time of approximately $160$ s, as can be seen in in Fig.\ref{fig:rate_errors}b. The associated mean relative pointing error (RPE) is found to be smaller than ${\rm 1\,\mu rad}$ in a 15 s time window corresponding to one experimental cycle.
Note that the spacecraft rotation rates grow to intolerably large levels of several hundred $\mu {\rm rad/s}$, if the spacecraft is freely floating without control loops to compensate the environmental torques (see Fig.\ref{fig:rate_errors}a). For orbit maintenance and de-orbiting, use of the CPS is foreseen.

\subsection{Micro-vibrations}
\begin{figure}
\begin{center}
\includegraphics[width=1.00\textwidth]{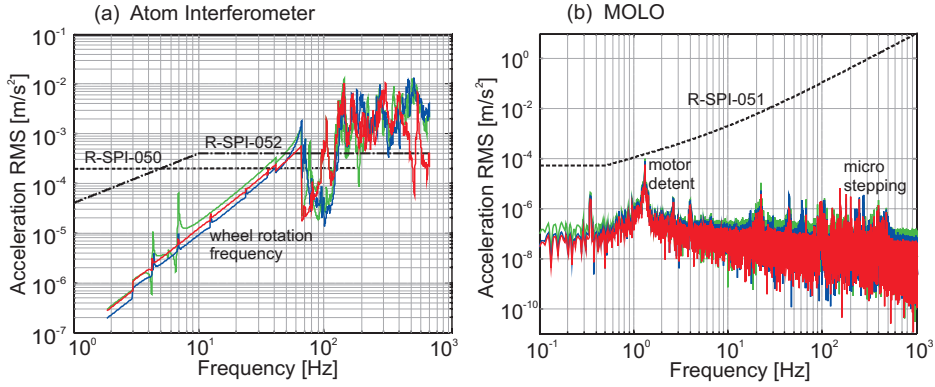}
\caption{Micro-vibrations induced by reaction wheels on the ATI-interface (left) and by the solar array driving mechanism on the MOLO interface for a SADM speed of 0.013 deg/s (right). Accelerations in x-direction (green), y-direction (blue), z-direction (red). Instrument requirements are represented by the dashed and dashed-dotted lines. The units used in the plot are RMS (root-mean-square) values of the acceleration ${\rm [ms^{-2}]}$. Note that for ease of comparison, requirements for the spectral density have been transformed into RMS requirements by integrating over the measurement bandwidth.}
\label{fig:micro_vibrations}
\end{center}
\end{figure}

The problems associated with micro-vibrations lie at the heart of two key trades for our current spacecraft configuration. One major cause of micro-vibrations are reaction wheels. To investigate the option of using them instead of the MPS for attitude control, extensive simulations were performed with a finite element model of the spacecraft which provided the transfer functions from the reaction wheels to the atom interferometer interface. Based on this model and available measurement data of low-noise reaction wheels (Bradford W18E), the level of micro-vibrations seen at instrument interface could be compared against the requirements. The results of the analyses are summarized in Fig. \ref{fig:micro_vibrations} left which clearly shows a violation of the requirement for frequencies above 20 Hz within the measurement band [1 mHz, 100 Hz]. The major contribution to the micro-vibrations comes from the H1 harmonic when the disturbance frequency corresponds to the wheel rotation rate up to the maximum wheel rate of 67 Hz. Further micro-vibrations are induced by the amplification of solar array flexible modes through the wheel rotation.
The violation of the requirement for the maximum level of tolerable micro-vibrations when using reaction-wheels, as found from our simulations, lead us to use a micro-propulsion system for attitude control in our baseline configuration.

Similar analyses were performed to determine the micro-vibrations induced by the solar array driving mechanism (SADM) on the MOLO interface, where a complex model of the SADM stepper motor together with the spacecraft finite-element-model (FEM) with deployed solar arrays was combined in a Matlab Simulink environment. The stepper motor generates two kinds of torque perturbations, one relating to the magnetic detent effect and the other relating to the commanded micro-stepping of the motor.
One particular resonance of the spacecraft structural model, corresponding to a torsional mode of the solar arrays around their rotation axis (y-axis), was found to be the most critical one.
When this resonance was hit by a harmonic disturbance originating from the detent torque of the solar arrays driving mechanism, the resulting response displayed a low-frequency isolated resonance peak at the excitation frequency of 1.3 Hz, with a few more peaks at higher frequencies due to harmonic excitation from the micro-stepping (see Fig. \ref{fig:micro_vibrations} right). Restricting the solar array rotation rates to below 0.13 deg/s, the level of micro-vibrations is found to be fully compliant throughout the specified frequency range, as can be seen in figure \ref{fig:micro_vibrations}. Allowing for higher rates up to 0.08 deg/s, which are required to follow the fast spacecraft rotations with the solar array at low altitudes below 10000 km, we find that only one specific rotation rate at 0.044 deg/s leads to a violation of the requirement. However, this is easily mitigated by quickly moving over the resonance, therefore avoiding to rotate the solar arrays at this particular rate.

\subsection{Science Links and Precise Orbit Determination}
For earth gravitational red-shift measurements of type (a), the measured frequency shift of the space-clock with respect to a ground clock is compared to the predicted frequency-shift. In order to compute the theoretical prediction, it is necessary to know the precise position and velocity of the space-clock (and therefore the precise orbit), as both quantities enter the equation describing the de-phasing of the space-clock proper time $\tau$ with respect to the coordinate time $t$, where the latter is referenced to an inertial earth-centered coordinate system. This dependence is expressed in Eq.\ref{equ:proper_time}, which is derived from a post-Newtonian expansion to second order in $1/c$ \cite{duchayne2009orbit}.
\begin{equation}
\label{equ:proper_time}
\frac{d\tau}{dt}=1-\left(\frac{U(\vec{x})}{c^2}+\frac{v^2}{2c^2}\right)+O({\rm c^{-4}}),
\end{equation}
where $U(\vec{x})$ denotes the gravitational potential, $\vec{x}$ the spacecraft position, and $v$ its velocity. Note that the first term in brackets corresponds to the gravitational time dilation whereas the second term corresponds to the special relativistic time dilation.
\\[1em]
{\it Clock and link uncertainty:} Considering only the gravitational time dilation, the frequency shift $\Delta\nu$ between a space clock at apogee and a ground clock may then be written as $\Delta\nu/\nu=[U(r_{\rm re})-U(r_{\rm ap})]/c^2=6.2\times10^{-10}$, where $r_{\rm re}$ and $r_{\rm ap}$ are the earth radius and the radial distance to apogee, respectively. It follows that the space-clock and the science links must have a fractional frequency uncertainty of better than $10^{-16}$ to allow measuring the gravitational red-shift of the earth to the specified accuracy. From similar deliberations one finds that the ground clock and science link performance must be even better, on the order of $10^{-18}$, to measure the gravitational red-shift of the sun to the specified accuracy. This requires very long contact times between the spacecraft and ground terminals, on the order of $2\times10^{5}\,{\rm s}$, which is only achieved after integration over many orbits.
\\The above requirements can be directly translated into the required stability for the link, expressed in modified Allan deviation\cite{allan1987time}, and into the frequency uncertainty for a space-to-ground or ground-to-ground comparison. A detailed performance error budget was established to assess all contributions to potential instabilities. For the microwave links we found that, provided the challenging requirements for high thermal stability are met, the ACES design with the suggested improvements can meet the specifications.
The assessment of the optical links showed that a significantly better performance than the state-of-the-art is required. A major issue for further investigation is the validity of extrapolating noise power spectral densities over long integration times in presence of atmospheric turbulence, which is required for comparing the estimated link performance to the specified link performance expressed in modified Allan deviation.
\\[1em]
{\it Orbit uncertainty:} As the fractional frequency uncertainty is specified to be around $10^{-16}$ for the space-clock, a target value of $3\times 10^{-17}$ for the orbit uncertainty does not compromise the overall measurement accuracy. Although preliminarily a constant target value of 2 m in position and 0.2 mm/s in velocity accuracy has been derived in \cite{cacciapuoti2013ste-quest}, these values can be significantly relaxed at apogee\cite{hechenblaikner2013gnss}. A final analysis will have to build on a relativistic framework for clock comparison as outlined in \cite{duchayne2009orbit}.
\begin{figure}
\begin{center}
\includegraphics[width=0.65\textwidth]{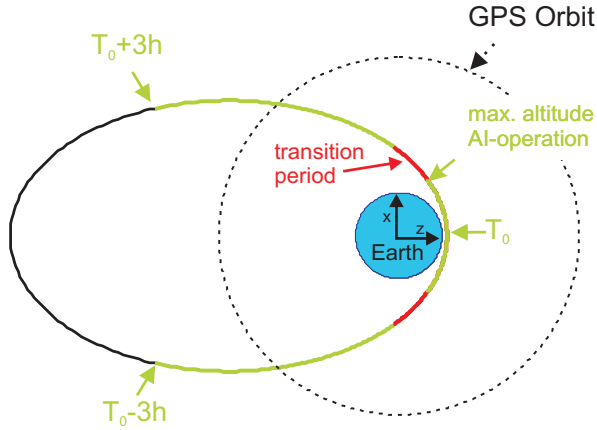}
\caption{The highly eccentric STE-QUEST orbit (solid line) and a GPS reference orbit (broken line). The orbit sections, where GNSS satellites may be tracked with sufficient accuracy, are plotted in green.}
\label{fig:gnss_visibility}
\end{center}
\end{figure}
\\[1em]
{\it GNSS Tracking:} The proposed method of Precise Orbit Determination (POD) relies on tracking  GPS and Galileo satellites with a GNSS receiver supporting
all available modulation schemes, e.g. BPSK5, BPSK10 and MBOC\cite{hofmann2008gnss}, and an antenna accommodated on the nadir panel. Considering that GNSS satellites broadcast their signal towards the earth and not into space, they become increasingly difficult to track at altitudes above their constellation (20200 km altitude), when the spacecraft can only receive signals from GNSS satellites behind the earth. In order to increase the angular range of signal reception, the receiver is required to support signal side-lobe tracking. This way, a good signal visibility for six hours during each orbit and for altitudes up to 34400 km can be achieved, with a corresponding PVT (position-velocity-time) error of 1.7 m at 1 Hz output rate\cite{hechenblaikner2013gnss}.
The transition maneuvers split the visibility region (green arch in Fig. \ref{fig:gnss_visibility}) into separate segments, which must be considered separately in the orbit determination. For both regions we find orbit errors less than 10 cm in position and less than 0.2 mm/s in velocity. Extrapolation of the spacecraft trajectory from the end of the visibility arch as far as apogee introduces large orbit errors ($\sim 10\,{\rm m}$) due to large uncertainties in parameters describing the solar radiation pressure. Nonetheless, these errors are sufficiently small to achieve the specified frequency uncertainty for the space-clock measurement if one considers the arguments for relativistic clock comparison derived in \cite{duchayne2009orbit}. A brief discussion of the possibility to relax the STE-QUEST orbit accuracy close to apogee is also given in \cite{hechenblaikner2013gnss}.

Finally, it shall be pointed out, that the accuracies obtained from GNSS-based POD are sufficient to probe the flyby anomaly, an inexplicable momentum transfer which has been observed for several spacecraft careening around the earth. These investigations can be performed without further modifications to the baseline hardware or mission concept, as described in Ref. \cite{Par2013}.

\section{Spacecraft subsystem design aspects}
\begin{figure}
\begin{center}
\includegraphics[width=1.00\textwidth]{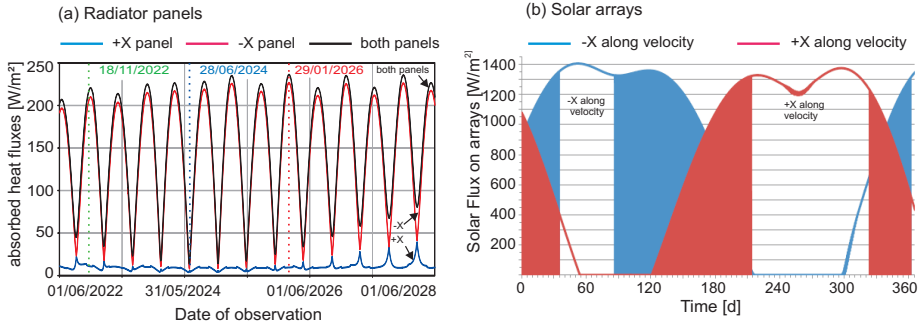}
\caption{(Left) The absorbed solar flux incident on the +X/-X radiator panels is plotted against the mission time. The dates for maximum temperature swing (green), hot case (red) and cold case (blue) are given by the respective dashed vertical lines. (Right) The solar flux incident on the solar arrays is plotted against time for one year. The blue and the red curves dare plotted for the X-panel along and opposite the spacecraft velocity, respectively.}
\label{fig:solar_flux}
\end{center}
\end{figure}

The major focus of the assessment study was to define preliminary designs of the spacecraft subsystems in compliance with all mission requirements. To this end, key aspects such as pointing strategy of spacecraft and solar arrays, variability of solar flux and temperature stability, measurement performance and pointing stability, instrument accommodation and radiative shielding, and launcher specifications, among others, must be reflected in the custom-tailored design features. Additional provisions for ease of access during integration and support of ground-based instrument testing on spacecraft level are also important considerations. As we intend to give a concise overview in this paper, we will limit our discussion on a few distinctive design features.
\\[1em]
{\it Thermal control subsystem (TCS) :} The STE-QUEST payload dissipates a large amount of power which may total up to more than 1.7 kW (including maturity and system margins) if all payload equipment is active. This represents a major challenge for the thermal system which can only be met through dedicated heat-pipes transporting the heat from the protected accommodation region in the spacecraft center to the radiator panels.  Also problematic is the fact that the baseline orbit is not sun-synchronous and features a drift of the right-ascension of the ascending node (RAAN), which leads to seasonally strongly variable thermal fluxes incident on the spacecraft from all sides. This is mitigated by optimal placement of the radiators in combination with a seasonal yaw-rotation by $180^{\circ}$ to minimize the variability of the external flux on the main radiators. The variation of the thermal flux during the 6 years of mission is plotted in Fig.\ref{fig:solar_flux} (left). The time when the largest temperature fluctuations occur during one orbit, defined as the maximum temperature swing, is close to the start of the mission and coincides with the time of the longest eclipse period. The additional use of heaters in proximity to the various payload units allows us to achieve temperature fluctuations of the payload units of less than $\pm2^{\circ}$C, whilst average operating points consistently lie within the required range from $10^{\circ}$C to $30^{\circ}$C.
During periods when only a minimum of solar flux is absorbed by the spacecraft, e.g. cold case of Fig.\ref{fig:solar_flux}, the heaters of the thermal control system may additionally contribute up to 150 W to the power dissipation.
These results were found on the basis of a detailed thermal model of the spacecraft implemented with ESATAN-TMS, a standard European thermal analysis tool for space systems\cite{ESATAN2010}.
\\[1em]
{\it Electrical Subsystem \& Communications:} Similar to the TCS, the design of the electrical subsystem is driven by the highly variable orbit featuring a large number of eclipses  (more than 1800 during the mission with a maximum period of 66 minutes), the high power demand of the instruments, and the satellite pointing strategy in addition to the required stability of the spacecraft power bus. In order to meet the power requirements, two deployable solar arrays at a cant angle of 45 degrees are rotated around the spacecraft y-axis. This configuration, in combination with two yaw-flips per year, ensures a minimum solar flux of approximately $1\,{\rm kW/m^2}$ on the solar panels (see Fig.\ref{fig:solar_flux} right), which generates a total power of 2.4 kW, including margins. The option of using a 2-axis SADM has also been investigated in detail and could be easily implemented. However, it is currently considered too risky as such mechanisms have not yet been developed and qualified in Europe and they might cause problems in relation to micro-vibrations.

As far as telemetry and tele-commanding (TT\&C) is concerned, contact times to a Northern ground station (Cebreros) are close to 15 h per day on average for the baseline orbit, which is easily compliant with the required minimum of 2 h per day. This provides plenty of margin to download an estimated 4.4 Gbit of data using a switched medium gain antenna (MGA) and low gain antenna (LGA) X-band architecture which is based on heritage from the LISA Pathfinder mission\cite{mcnamara2008lisa}. The switched TT\&C architecture allows achieving the required data volume at apogee on the one hand, whilst avoiding excessive power flux at perigee to remain compliant with ITU regulations on the other hand.
\\[1em]
{\it Mechanical subsystem, assembly integration and test (AIT) and radiation aspects:}
The spacecraft structure is made almost entirely from panels with an aluminum honeycomb structure (thickness 40 mm) and carbon-fibre reinforced polymer (CFRP) face sheets, which is favorable from a mechanical and mass-savings perspective.
A detailed analysis based on a finite-element-model (FEM) of the spacecraft structure revealed a minimum eigenfrequency of 24 Hz in transverse and 56 Hz in axial direction, well above the launcher requirements of 15 Hz and 35 Hz, respectively. The first eigenmode of the atom interferometer physics package, corresponding to a pendulum mode in radial direction, was found at 67 Hz.
The spacecraft structure is designed to optimize AIT and aspects of parallel integration. To this end, the two instruments can be integrated and tested separately before being joined on payload module level without the need to subsequently remove any unit or harness connection. The heat-pipe routing through the spacecraft structure allows functional testing of both instruments on ground by operating the heat-pipes either in nominal mode or re-flux mode (opposite to the direction of gravity). As an important feature in the integration process, payload and service module components are completely separated in their respective modules up to the last integration step, when they are finally joined and their respective harness connected on easily accessible interface brackets.

The second major aspect in structural design and instrument accommodation has been to minimize radiation doses of sensitive instrument components.  As the spacecraft crosses the van Allen belt twice per orbit and more than 5000 times during the mission, it sustains large radiation doses from trapped electrons and solar protons.
Dedicated design provisions and protected harness routing ensure that the total ionizing doses (TIDs) are below 30 krad for all instrument units and below 5 krad for the sensitive optical fibres. The non-ionizing energy loss, which is primarily caused by low-energy protons, is found to be well below $2\times10^{10}\,{\rm m^{-2}}$, expressed in terms of 10 MeV equivalent proton fluence. These results were obtained from the total dose curves accumulated over the orbit during the mission in combination with a dedicated sectoring analysis performed with a finite-element-model of the spacecraft.

\section{Conclusions}
STE-QUEST aims to probe the foundations of Einstein's Equivalence Principle by performing measurements to test its three cornerstones, i.e. the local position-invariance, the local Lorentz-invariance and the weak equivalence principle, in a combined mission. It complements other missions which were devised to explore the realm of gravity in different ways, including the soon-to-be-launched ACES\cite{salomon2001cold} and MICROSCOPE\cite{Touboul2001a} missions and the proposed STEP\cite{sumner2007step} mission.

The STE-QUEST measurements are supported by two instruments.
The first instrument, the atomic clock, benefits from extensive heritage from the ACES mission\cite{salomon2001cold}, which therefore reduces associated implementation risks. The other instrument, the atom interferometer, would be the first instrument of this type in space and poses a major challenge which is currently met by dedicated development and qualification programs.
The mission assessment activities summarized in this paper yielded a spacecraft and mission design that is compliant with the challenging demands made by the payload on performance and resources.

Several critical issues have been identified, including low maturity of certain payload components, potential unavailability of sufficiently accurate ground clocks, high sensitivity of the optical links to atmospheric distortions and associated performance degradation, and high energy dissipation of the instruments in addition to challenging temperature stability requirements. However, none of these problems seems unsurmountable, and appropriate mitigation actions are already in place.

\begin{acknowledgements}
The work underlying this paper was performed during the mission assessment and definition activities (Phase 0/A) for the European Space Agency (ESA) under contract number 4000105368/12/NL/HB.
The authors gratefully acknowledge fruitful discussions and inputs from the ESA study team members, in particular Martin Gehler (study manager), Luigi Cacciapuoti (project scientist), Robin Biesbroek (system engineer), Astrid Heske, (payload manager), Florian Renk (mission analyst), Pierre Waller, (atomic clock support), and Eric Wille (atom interferometer support). We also thank the Astrium team members for their contributions to the study, in particular Felix Beck, Marcel Berger, Christopher Chetwood, Albert Falke, Jens-Oliver Fischer, Jean-Jacques Floch, S\"oren Hennecke, Fabian Hufgard, G\"unter Hummel, Christian Jentsch, Andreas Karl, Johannes Kehrer, Arnd Kolkmeier, Michael G. Lang, Johannes Loehr, Marc Maschmann, Mark Millinger, Dirk Papendorf, Raphael Naire, Tanja Nemetzade, Bernhard Specht, Francis Soualle and Michael Williams. The authors are grateful for important contributions from our project partners Mathias Lezius (Menlo Systems), Wolfgang Sch\"afer, Thorsten Feldmann  (TimeTech), and Sven Sch\"aff (Astos Solutions). Finally, we are indebted to R\"udiger Gerndt and Ulrich Johann (Astrium) for regular support and many useful discussions.
\end{acknowledgements}


\end{document}